\documentclass[useAMS,usenatbib]{biom}

\providecommand{\normaldistn}{\mathrm{Normal}}  
\providecommand{\prob}{\mathrm{P}}              
\providecommand{\uisd}{\sigma}                  
\providecommand{\yt}{\bar{y}_{\mathrm{T}}}      
\providecommand{\yc}{\bar{y}_{\mathrm{C}}}      
\providecommand{\yx}{\bar{y}_{\mathrm{E}}}      
\providecommand{\sigt}{s_{\mathrm{T}}}          
\providecommand{\sigc}{s_{\mathrm{C}}}          
\providecommand{\sigx}{s_{\mathrm{E}}}          
\providecommand{\nt}{n_{\mathrm{T}}}            
\providecommand{\nc}{n_{\mathrm{C}}}            
\providecommand{\nx}{n_{\mathrm{E}}}            
\providecommand{\thetat}{\theta_{\mathrm{T}}}   
\providecommand{\thetac}{\theta_{\mathrm{C}}}   
\providecommand{\thetax}{\theta_{\mathrm{E}}}   

\providecommand{\muc}{\mu_{\mathrm{C}}}         

\usepackage[figuresright]{rotating}
\usepackage{url}
\usepackage{color}

\title[Discussion: ``Principled type~I error rate inflation\ldots'']{Discussion on ``Principled type~I error rate inflation \\ in two-arm clinical trial designs with external control information borrowing''\\ by Silvia Calderazzo, Manuel Wiesenfarth, Vivienn Weru,\\ and Annette Kopp-Schneider}

\author{Tim Friede$^{*}$\email{tim.friede@med.uni-goettingen.de} \\
        Department of Medical Statistics,\\ University Medical Center G\"{o}ttingen,\\ Humboldtallee~32, 37073~G\"{o}ttingen, Germany
        \and 
        Christian R\"{o}ver\\
        Department of Medical Statistics,\\ University Medical Center G\"{o}ttingen,\\ Humboldtallee~32, 37073~G\"{o}ttingen, Germany
	   }

\begin{document}


\pagerange{\pageref{firstpage}--\pageref{lastpage}} \pubyear{2026}

\artmonth{October}


\label{firstpage}


\begin{abstract}
  Consideration of frequentist \mbox{type-I} error or power based on certain fixed conditions may not fully embrace the aims of a Bayesian analysis; this is particularly true in settings where informative priors are involved.
  Calderazzo \emph{et~al.} consider adaptations of a testing procedure involving Bayesian borrowing from external control data.
  Particular caution may be necessary here in order not to condition on parameters and data simultaneously.
  We consider their example case and argue that (a)~the conditioning scheme may have limited practical implications, and (b)~if the main concern are assumption violations, these might be better addressed by adapting the likelihood function. 
\end{abstract}

%
%


\maketitle

\section{Introduction}\label{sec:intro}
  We congratulate \citet*{CalderazzoEtAl2026} for their intriguing and unconventional approach to borrowing from external data while bounding the eventual \mbox{type-I} error.
  Their proposal provides an interesting angle on the problem, but we would also like to highlight some concerns that we have with the approach.

  Once Bayesian methods or aspects are implemented in a statistical model, classical ``frequentist'' operating characteristics (unbiasedness, \mbox{type-I} error, coverage, based on fixed parameter values) may not necessarily be suitable.
  This holds in particular once informative prior distributions are involved, which are often utilized to implement \emph{borrowing} from external information \citep{LesaffreEtAl2024}.
  Within their power or \mbox{type-I} error considerations, \citet{CalderazzoEtAl2026} rely on marginalization over the experimental data while fixing (or \emph{conditioning on}) certain figures.
  We question whether the conditioning scheme employed is suitable for the intended purpose, and what exact kind of scenario is targeted here.
  If model misspecification is the primary concern, \citet{BestEtAl2025} have suggested alternative strategies to gauging the performance of statistical methods.
  Also, if the aim is to safeguard against assumption violations, then this might be better addressed by adapting the likelihood specification rather than decision thresholds.

\section{The example setup}\label{sec:examplesetup}
  In the following, we present the model considered by \citet[Sec.~2]{CalderazzoEtAl2026} using a different, but equivalent notation.
  The main difference is that they phrased the external control mean in terms of a prior distribution, whereas we explicitly consider the external controls as data.
  We consider treatment and control data from samples of sizes~$\nt$ and~$\nc$, respectively.
  From both samples, we observe averages~$\yt$ and~$\yc$ 
  with associated standard errors $\sigt=\frac{\uisd}{\sqrt{\nt}}$ 
  and $\sigc=\frac{\uisd}{\sqrt{\nc}}$.
  In addition, we have external (control) data~$\yx$ (of size~$\nx$) 
  with standard error $\sigx=\frac{\uisd}{\sqrt{\nx}}$.
  The external data are relevant to the problem because they provide additional evidence for the control group's mean.
  This is implemented via the likelihood:
  \begin{eqnarray}
    \yt|\thetat,\sigt &\sim& \normaldistn\bigl(\thetat,\, \sigt^2\bigr) \mbox{,} \label{eqn:likeli01} \\
    \yc|\thetac,\sigc &\sim& \normaldistn\bigl(\thetac,\, \sigc^2\bigr)\mbox{,} \label{eqn:likeli02}\\
    \mbox{and} \quad \yx|\thetac,\sigx &\sim& \normaldistn\bigl(\thetac,\, \sigx^2\bigr) \mbox{.} \label{eqn:likeli03}
  \end{eqnarray}
  Note that control and external data share a common mean parameter~($\thetac$), 
  while it differs for the treatment data~($\thetat$).
  
  Ultimately, interest is in inferring the contrast of treatment and control parameters ($\delta=\thetat-\thetac$).
  Aiming for a non-informative specification, (improper) uniform priors for both~$\thetat$ and~$\thetac$ are assumed.
  In case the external data were available \emph{before} the main experiment, one might actually proceed in two stages. $\yx$~may be analyzed first, yielding the external-data posterior
  \begin{eqnarray}
    \thetac | \yx, \sigx & \sim & \normaldistn\bigl(\yx,\, \sigx^2\bigr) \label{eqn:infoprior}
  \end{eqnarray}
  \citep{BDA3rd}.
  Subsequently, the control data~($\yc$) may then be analyzed using the above posterior as the (informative) prior.
  Note that this prior's mean~($\yx$) was denoted as~$\muc$ in \citet{CalderazzoEtAl2026}, and $\muc$~was primarily considered as part of the prior specification, rather than part of the data.

\section{The data model as a DAG}
\begin{figure}[b]
  \begin{center}
    \includegraphics[width=0.9\linewidth]{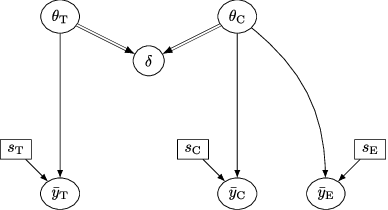}
  \end{center}
  \caption{The model from Section~\ref{sec:examplesetup} illustrated in terms of a \emph{directed acyclic graph (DAG)}
  \citep{GreenlandPearl2010}.
  The likelihood expressions~(\ref{eqn:likeli01})--(\ref{eqn:likeli03}) are reflected by terminal nodes ($\yt$, $\yc$ and $\yx$) whose parental nodes are the mean parameters ($\thetat$ or $\thetac$, which may be estimated from the data) and standard errors ($\sigt$, $\sigc$ or $\sigx$, which are treated as constants here). The treatment effect~$\delta$ is a deterministic function of the two mean parameters ($\delta=\thetat-\thetac$), which is indicated by a double arrow.
  \label{fig:dag}}
\end{figure}

  This model may also be illustrated in terms of a \emph{directed acyclic graph (DAG)} 
  \citep{Pearl1995,GreenlandPearl2010}; 
  a corresponding DAG is shown in Figure~\ref{fig:dag}. 
  The advantage of the DAG representation is that it provides a visualisation of the dependence of parameters and observables.
  For example, when fixing (conditioning on) the value of~$\thetac$, one can read off from the DAG that~$\yc$ and $\yx$ are \emph{conditionally independent given~$\thetac$}, since the only connection between~$\yc$ and~$\yx$ is via~$\thetac$. The two (conditional) distributions of~$\yc$ and~$\yx$ are given in~(\ref{eqn:likeli02}) and~(\ref{eqn:likeli03}).
  On the other hand, when conditioning on~$\yx$, the conditional distribution of~$\yc$ depends on the value of~$\yx$ due to the connection via~$\thetac$. We have already come across the conditional distribution~$\prob(\thetac|\yx)$ in~(\ref{eqn:infoprior}), while the conditional distribution~$\prob(\yc|\thetac)$ was given in~(\ref{eqn:likeli02}); together, these imply that
  \begin{eqnarray}\label{eqn:conditional}
    \yc | \yx, \sigx, \sigc & \sim & \normaldistn\bigl(\yx,\, \sigx^2+\sigc^2\bigr) \mbox{.}
  \end{eqnarray}

\section{Conditioning}
  As we have seen in Figure~\ref{fig:dag}, one may condition on different nodes in the graph.
  Conditioning on the parameters ($\thetac$~and~$\thetat$, or $\thetac$~and~$\delta$) yields joint conditional distributions of data ($\yt$, $\yc$, $\yx$) in a certain scenario regarding the ``true parameters''.
  Conditioning on the data ($\yt$, $\yc$ and $\yx$) yields the joint posterior distribution of the parameters $\thetat$ and $\thetac$ (and with that, of~$\delta$).

  One may consider \emph{conditioning} as posing a certain \emph{what if} question (such as, ``what if $\delta=0\,$?'', when interested in \mbox{type-I} error computations).
  In a Bayesian context, conditioning on certain parameters may for example be useful for prior predictive checks. Most importantly, conditioning on the data yields the posterior distribution that is the basis for inference.
  Conditioning on certain parameter values and considering the coverage probabilities of inferences based on posterior intervals will in general not yield coverages according to the nominal level; this is only guaranteed on average over the prior (provided that a proper prior was specified) \citep{CookGelmanRubin2006}.

  The power / \mbox{type-I} error~$\beta^\pi(\thetac,\thetat)$ considered by \citet[Sec.~2]{CalderazzoEtAl2026} is in fact a function of $\thetac$, $\thetat$ \emph{and}~$\yx$, so that implicitly this corresponds to \emph{conditioning on all three}.
  This is not an uncommon approach, and has previously been considered by others in a similar manner \citep[e.g.,][]{VieleEtAl2014,SchmidliEtAl2014,Weber2026}.
  While one may freely condition on~$\thetat$ and~$\thetac$ (or~$\thetac$ and~$\delta$) without ever challenging the model assumptions (also because improper, location-invariant priors were specified), this is trickier with other figures that are connected via the graph.
  For example, with the settings of~$\sigc=0.22$ and~$\sigx=0.32$, the expression in~(\ref{eqn:conditional}) implies a variance of~$0.39^2$;
  any scenario with $|\yc-\yx| > 2\sqrt{\sigc^2+\sigx^2} = 0.77$ would hence tend to be unlikely or suspicious --- and may imply a discrepancy close to the anticipated treatment effect~$\delta$.
  
  It is not clear what exact question is addressed by considering marginalization over data~$\yc$ while conditioning on both~$\thetac$ and~$\yx$.
  Considering the \emph{long-run average} motivation commonly utilized in frequentist arguments, this would seem to correspond to pondering a universe in which for a fixed parameter~$\thetac$ the external data~$\yx$ would remain constant while the concurrent control data~$\yc$ would still vary --- in which case the the model would clearly be wrong.
  Within the model as originally stated, it does not make sense to independently fix~$\thetac$ and~$\yx$; the model implies a close relationship between the two, as exemplified in (\ref{eqn:likeli03}) and~(\ref{eqn:infoprior}).
  By conditioning on both~$\thetac$ and~$\yx$, this relationship appears to be ``hot-wired''; plugging in two fixed numbers may not be helpful for assessing the stochastic nature or their relationship.
  If only $\thetac$~was fixed, then $\yx$ would vary in unison with~$\yc$. If only $\yx$~was fixed, then $\thetac$ could not take arbitrary values without essentially violating the model; it would reasonably only occupy a limited range defined through the corresponding conditional~(\ref{eqn:infoprior}).
  It remains unclear whether the distribution averaged over is meant to explicitly challenge the model assumptions, or what kind of model violations are meant to be scrutinized here.

\begin{figure}[b]
  \begin{center}
    \includegraphics[width=0.9\linewidth]{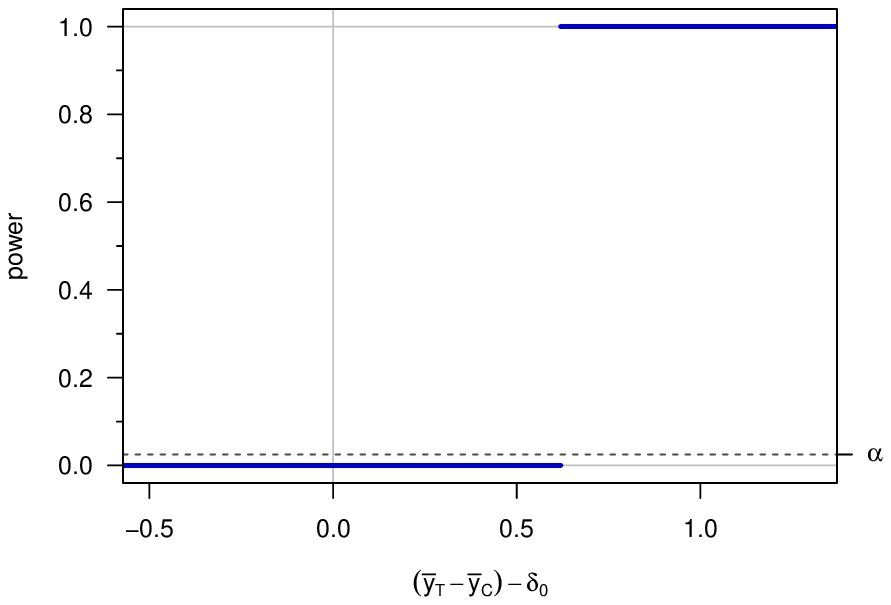}
  \end{center}
  \caption{A power curve (shown in blue) according to equation~(\ref{eqn:powercurve}),
  with $\sigt\!=\!\sigc\!=\!\frac{1}{\sqrt{20}}\!=\!0.22$ and~$\alpha\!=\!0.025$. 
  As a result from conditioning on both parameters \emph{and} data, 
  the power is either zero or one.
  The targeted \mbox{type-I} error (of~$\alpha\!=\!0.025$) is only achieved when averaging over the data-generating process.
  \label{fig:powercurve}}
\end{figure}

  The problem appears to originate from simultaneous conditioning on parameter(s) \emph{and} downstream data.
  An analogous argument may be made based on the plain frequentist test decision (ignoring the external data~$\yx$).
  The classical frequentist test would reject the null hypothesis ($H_0:\, \delta\leq\delta_0$) if
  \begin{eqnarray}
    \frac{(\yt-\yc)-\delta_0}{\sqrt{\sigt^2 + \sigc^2}} & > & z_{1-\alpha} \mbox{.}
  \end{eqnarray}
  We may now consider the corresponding \emph{power function} that conditions on \emph{both} parameter(s) \emph{and} data:
  \begin{eqnarray}\label{eqn:powercurve}
    &&\prob\biggl(\textstyle \frac{(\yt-\yc)-\delta_0}{\sqrt{\sigt^2 + \sigc^2}} > z_{1-\alpha} \;\bigg|\; \delta,\, \yt,\, \yc\biggr) \nonumber \\
    &=&
    \left\{ 
      \begin{array}{ll}
        1 & \mbox{if } \textstyle \yt -\yc > \delta_0 + z_{1-\alpha}\sqrt{\sigt^2+\sigc^2}\\
        0 & \mbox{otherwise.}\\
      \end{array} \right.
  \end{eqnarray}
  This power expression based on fixed parameter \emph{and} data is either zero or one, and in particular, it is nowhere near the targeted level of~$\alpha$.
  A corresponding power curve is illustrated in Figure~\ref{fig:powercurve}.
  
  Again, drawing upon the \emph{long-run average} motivation, these power values of zero and one relate to scenarios in which the likelihood assumption would essentially be wrong.
  Once historical data are included in the set-up, the effect may be less drastic, but the issue remains the same: if parameters \emph{and} data are conditioned upon, deviations from the targeted power level are natural and not necessarily a cause for concern.
  
  If one wanted to safeguard against model violations, it is more instructive to explicitly specify a corresponding alternative data-generating process (likelihood) and consider operating characteristics under such an alternative. This would in fact be a very constructive way of scrutinizing a model, as an alternative likelihood may immediately be adopted for inference, if it is in fact considered more plausible (or robust, or conservative);
  such an approach has previously also been advocated by \citet{BestEtAl2025}.
  In the present case, one might for example introduce a separate parameter~$\thetax$ for the external data along with assumptions on the (dis-)similarity to the control parameter~$\thetac$.
  
\section{Discussion}\label{sec:discuss}
  From a Bayesian viewpoint, considerations of classical frequentist \mbox{type-I} error probabilities based on singular fixed parameter values often seem to be a distraction, in particular, when dealing with bounded parameter spaces or informative prior distributions.
  \emph{Marginally}, on average over the prior distribution, proper calibration meanwhile is guaranteed by construction \citep{CookGelmanRubin2006}. 
  This is also acknowledged in the recent FDA draft guidance on the use of Bayesian methodology, which points out that
  ``direct interpretation of posterior probability [\ldots] can be appropriate in Bayesian analyses that explicitly leverage external data sources [\ldots].''
  \citep{FDA2026}
  In the present application, the inference problem may be phrased in terms of a simple two-group comparison with an informative prior for one of the group means, or, equivalently, the external information may be considered as part of the likelihood function.
  The marginalization scheme employed here to derive (conditional) \mbox{type-I} errors appears to be based on a rather artificial set-up,
  so that the suitability to evaluate and judge operating characteristics is questionable.

  As the overarching aim appears to be to accommodate potential prior/data conflict (assumption violations), we feel that this purpose may be served better by anticipating it in the model setup; this could for example be approached by including an additional variance (heterogeneity) parameter \citep{RoeverFriede2020,RoeverFriede2026}, by including reasonable alternative models in a \emph{model averaging} or \emph{robustification} approach \citep{RoeverWandelFriede2019,SchmidliEtAl2014}),
  or by utilizing heavy-tailed priors or likelihoods to prioritize concurrent over external data \citep{OHaganPericchi2012}.
  Similarly, if for example the assumption of a common within-group variance~($\uisd$) was considered doubtful, this would hardly be addressed by adjusting the significance level, but rather by accommodating separate variance parameters in the model.
  It seems odd that disagreement with the model assumptions would reasonably be addressed by tinkering with decision thresholds rather than with the likelihood itself.
  Also, instead of considering a hybrid frequentist/Bayesian setup, an instructive exercise might also be to implement the hierarchical model as outlined above in a fully frequentist fashion --- where conditioning on data and parameters does not seem to be common practice.

\backmatter


\section*{Acknowledgements}
This work was funded by the European Union's Horizon Europe Research and Innovation program under grant agreement 101136365 (INVENTS, \url{https://invents-he.eu/}), co-funded by the Swiss State Secretariat for Education, Research and Innovation (SERI) and by the UKRI Innovative UK under their Horizon Europe Guarantee scheme.
Views and opinions expressed are those of the authors only and do not necessarily reflect those of the European Union or the European Health and Digital Executive Agency. Neither the European Union nor the granting authority can be held responsible for them.
Tim Friede is grateful to University Medical Center G\"{o}ttingen for the opportunity of a sabbatical (Forschungssemester) in the summer term~2026.

\section*{Data availability}
Data sharing is not applicable to this article as no new data were created or analyzed.

%
\vspace*{-8pt}


\bibliographystyle{abbrvnat}
\bibliography{literature}

%

%

\label{lastpage}

\end{document}